# Engaging the Public with Supernova and Supernova Remnant Research Using Virtual Reality

**Gilles Ferrand**
RIKEN Cluster for Pioneering Research, Astrophysical Big Bang Laboratory (ABBL)
gilles.ferrand@riken.jp

**Don Warren**
RIKEN Interdisciplinary Theoretical and Mathematical Sciences Program (iTHEMS)
donald.warren@riken.jp

On 21 April 2018, the citizens of Wako, Japan, interacted in a novel way with research being carried out at the Astrophysical Big Bang Laboratory (ABBL) at RIKEN[1,2]. They were able to explore a model of a supernova and its remnant in an immersive three-dimentional format by using virtual reality (VR) technology. In this article, we explain how this experience was developed and delivered to the public, providing practical tips for and reflecting on the successful organisation of an event of this kind.

## Introduction

Visualisation is an essential part of research, both to explore one's data and to communicate one's findings with others. Many data products in astronomy come in the form of three-dimensional (3D) cubes, and since our brains are tuned for recognition in the 3D world, we ought to display these in 3D space to get a better idea of what is being observed. This kind of experience is possible with virtual reality (VR) devices[3].

VR technology has huge potential for education and public outreach. Objects vastly larger (galaxies) or smaller (nuclei) than the human visual range can be presented at an understandable scale in an immersive 3D space. VR can allow non-experts to more directly connect with the science that is presented, and being at the cutting edge of technology, VR naturally appeals to young audiences, which are arguably the most important audiences to consider for the future of our discipline. Another option to increase the appeal of scientific data among lay people is 3D printing (Arcand et al., 2017).

Alongside our research, we have been developing VR visualisations of our simulation data. These were first used with our colleagues, that is, other astrophysicists[4]. Then, during the internal centennial celebrations at RIKEN in December 2017, the visualisations were used to showcase the work done at the ABBL to other employees, most of whom were not research personnel (Figure 1). On receiving positive reactions from the attendees, our laboratory decided to set up a VR booth for the annual Open Day of RIKEN held on 21 April 2018 at the main campus, in Wako[5], Japan. In the following article, we describe our experience, describe how we visualised our data in immersive interactive 3D format and illustrate in detail the organisation and outcomes of the event.

## Experience

Our aim is to enhance scientific visualisation; to this end, we produced a visual rendering of actual science data. In our case, the data are simulated data, although the technique outlined in the next section applies equally well to observational data. Even though the visualisation is created on a computer and great attention has been paid to the way it looks, we are not showing an artist's rendition or computer-generated imagery (CGI). In an effort to engage the public with our research, we want to show them the actual output of our work.

Research at the Astrophysical Big Bang Laboratory (ABBL) focuses on explosive astrophysical phenomena and energetic particles. One of our current projects is to link studies on supernovae (SNe) and studies of supernova remnants (SNRs)[6]. Accordingly, our demo (dubbed SN2SNR) has two parts.

The first part shows data from simulations of a supernova, provided by our collaborator Fritz Röpke in Germany (published in Seitenzahl et al. 2013). It shows the structure of the supernova and the distribution imprinted on the SNR phase from the SN phase (compare the two rows of Figure 3). Studying this imprinting is the scientific goal of the research project.

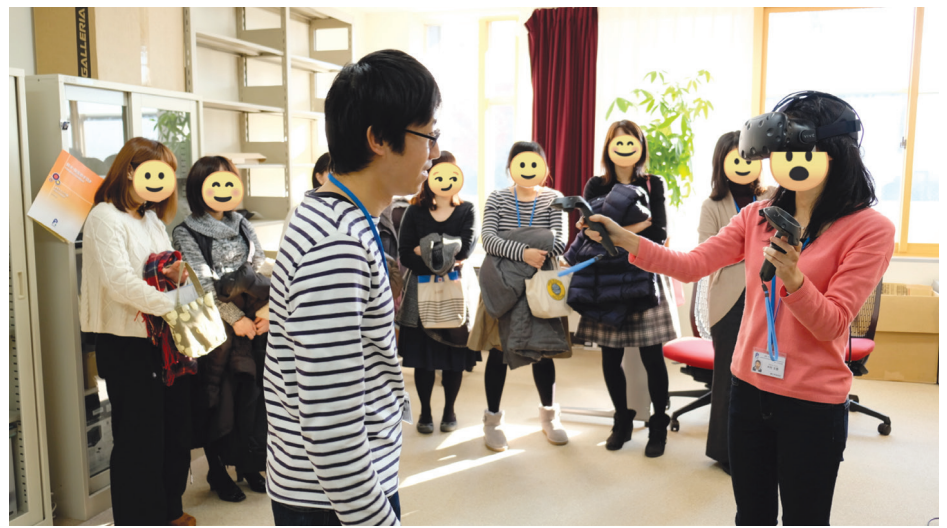

**Figure 1.** *Employees of the accounting department exploring a 3D model of a supernova at the ABBL during the RIKEN Centennial Meeting (faces masked for privacy). Credit: RIKEN/ABBL*

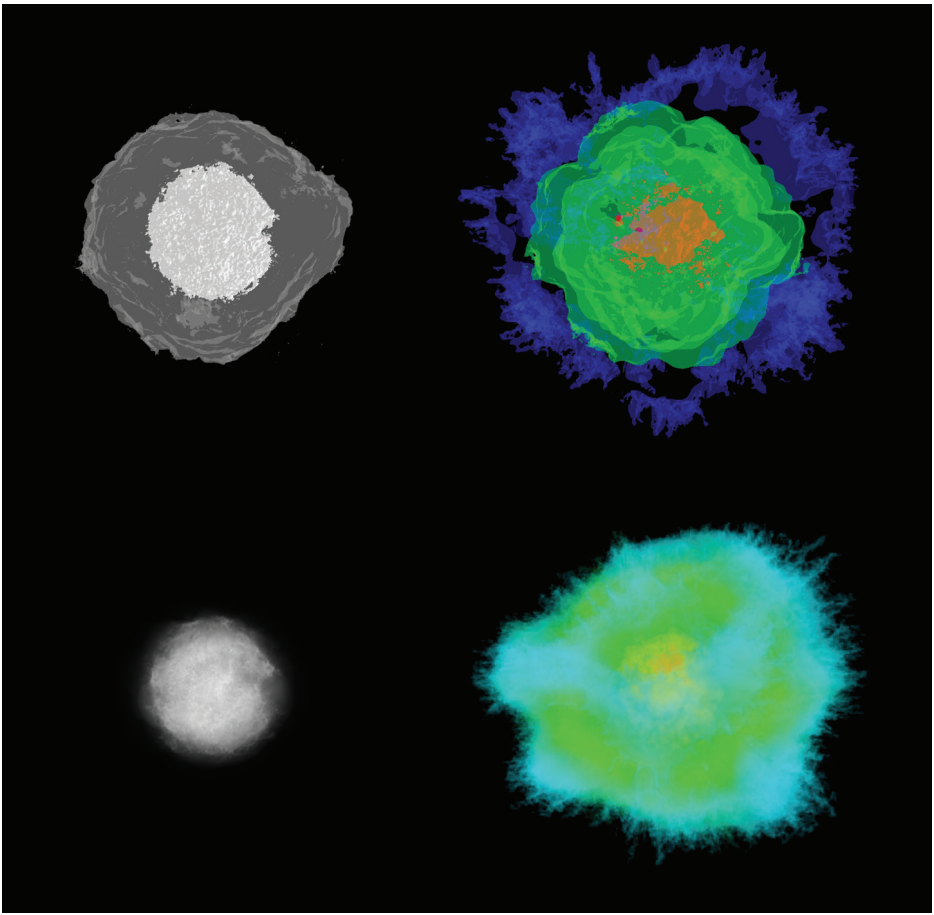

### Development: The Hardware and Software

The project presented here builds on expertise gained during a pilot project conducted at the University of Manitoba in 2016[7]. The technical aspects related to VR were developed by the authors at RIKEN over about a year from the spring of 2017 to that of 2018. Here, we discuss the tools we chose and the parts we had to develop ourselves. The general architecture of the system is shown in Figure 4.

**Figure 2.** *Different visualisations of a supernova explosion. Models on the left show the mass density, while models on the right show the abundances of three chemical elements after the computation of explosive nucleosynthesis (red: nickel-56 that will decay into iron; green: oxygen; blue: carbon). Models at the top show isosurfaces (at 0.1% and 10% of the maximum for the density, at 90% for the abundances), while models at the bottom show volume rendering of the entire data cube. Credit: RIKEN/ABBL*

of chemical elements. The way it looks on a computer screen is shown in Figure 2. Of course, these snapshots cannot really convey the 3D feel of it. This part of the demo also presents two different kinds of 3D visualisation: *isosurfaces*, where the set of points at a given data value are displayed rendered as a mesh (top of Figure 2), and *volume rendering*, where the entire data cube is displayed rendered as a glowing gas (bottom of Figure 2).

The second part of the demo shows data from simulations of the subsequent supernova remnant phase developed by Gilles Ferrand, wherein the output of the supernova simulation is the input for the SNR simulation. The research paper for this is in preparation but the general method is as in Ferrand et al. (2010). This part of the demo uses a custom volume rendering of the shell of shocked material that makes apparent the unstable boundary of the material ejected from the supernova. The way it looks on a computer screen is shown in Figure 3. The user can switch between two different initial conditions for the SNR phase in order to highlight what is

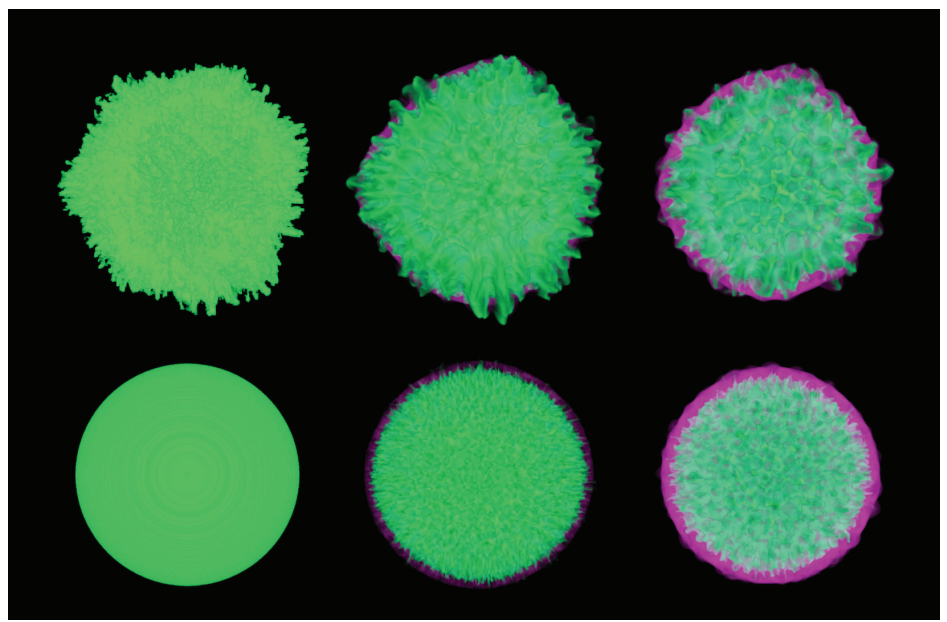

**Figure 3.** *Visualisation of the simulation of a supernova remnant. The quantity shows the mass density and the colour codes show the material: the stellar ejecta are green, while the interstellar medium (ISM) swept-up by the blast wave is purple. Only the shell of the shocked material is shown. The interface between the ejecta and the ISM is deformed because of Rayleigh–Taylor instability. The columns show the time evolution. Left: one year, centre: 100 years, right: 500 years. Images shown are on the same scale, and over this period of time, the SNR actually expands by a factor of about 150 (which is fine when rendered in VR). The two rows compare different initial conditions: turbulent at the top (as obtained from the SN simulation of Figure 2) and spherically symmetric at the bottom (after averaging out over angles). Credit: RIKEN/ABBL*

Different VR solutions are available, in widely different forms and with varying price ranges, from Google Cardboard[8] to the Visbox CAVE[9]. We chose the recently released HTC VIVE[10] headset, because we found that it offers great immersion with a relatively simple setup and is reasonably priced. The VIVE consists of a head-mounted stereoscopic display for creating images in 3D, two handheld controllers for in-world user interaction and two lighthouses which enable precise room-scale tracking of the headset and controllers. It is the combination of these that creates the full VR experience. It is important to stress that a convincing 3D model is not just about stereo vision. A 3D movie brings depth but is scripted and remains a passive experience seen from a pre-selected viewpoint. In VR, the user is able to actively explore the 3D content presented. Further, we do not merely seek a 360 degree experience, as is the case on smartphone-based systems, where one can only look around. The 6-degrees-of-freedom (d.o.f.) of the VIVE allow the user to freely move in the room around the virtual object and manipulate it as if it is really there.

On the software side, we rely on the Unity game engine[11]. We are not aware of a solution that provides what we want out-of-the-box, and rather than trying to extend an existing astronomy visualisation programme to advanced displays, we have taken the alternative route of customising one of the most popular 3D engines for game development[12]. While this may seem surprising at first, building on a high-level standard solution offers several advantages, in particular, fast prototyping and testing. Unity has already been used for a number of serious applications in the medical and architectural fields as well as in the natural sciences[13, 14]. Importantly, to us, it offers support for all existing VR displays, eliminating the burden of technical aspects like stereoscopy or tracking, so that we can concentrate on data rendering and the user interface (see below). For the VIVE, support is provided through the official SteamVR plugin[15]. The main disadvantage of this approach is that its versatility may result in suboptimal performance.

## Making it Work for Science

There were some technical aspects that had to be resolved in order to use Unity for astronomy (see also the earlier progress report Ferrand et al., 2016[16]).

The first step is to load the data. As volume rendering is done inside Unity, the raw data file had to be converted into a 3D texture format that could be loaded on to the graphics card. Isosurfaces, on the other hand, had to be extracted from the data cube beforehand (using visualisation software like VisIt[17] or Python/scikit-image[18]). Once saved as a standard mesh file (like in the .obj format), the data can be readily loaded inside Unity and placed in a 3D scene.

The next step is to create the desired visualisation. For volume rendering, we perform ray casting, where a colour and opacity are assigned to each volume element (voxel), and the RGBA colour is integrated along each line of sight (where R, G and B are the colour channels and A is the opacity). This is implemented using custom shaders (small programmes that run on the graphics processing unit), which gives us full control over the transfer function that maps data values into voxel colours. While ray-tracing is precise, it is computationally expensive, and this limits the data size we can handle to keep the user experience seamless[19]. For meshes representing isosurfaces, we need an external lighting source, so that they actually look three dimensional (the presence of shadows makes a considerable difference).

The final step is to code the interaction of the user with the demo. This involves mapping user inputs into the desired functionality. We offer manipulation of the cube with the controllers, either via the built-in trackpad (which users who have used a desktop setup will be familiar with) or with more native interactions: moving or rotating one's hands in space to impart geometrical transformations to the data cube. Offering a natural user experience is important, both so that researchers can be productive and so that public visitors engage well with it. We also localised the user interface

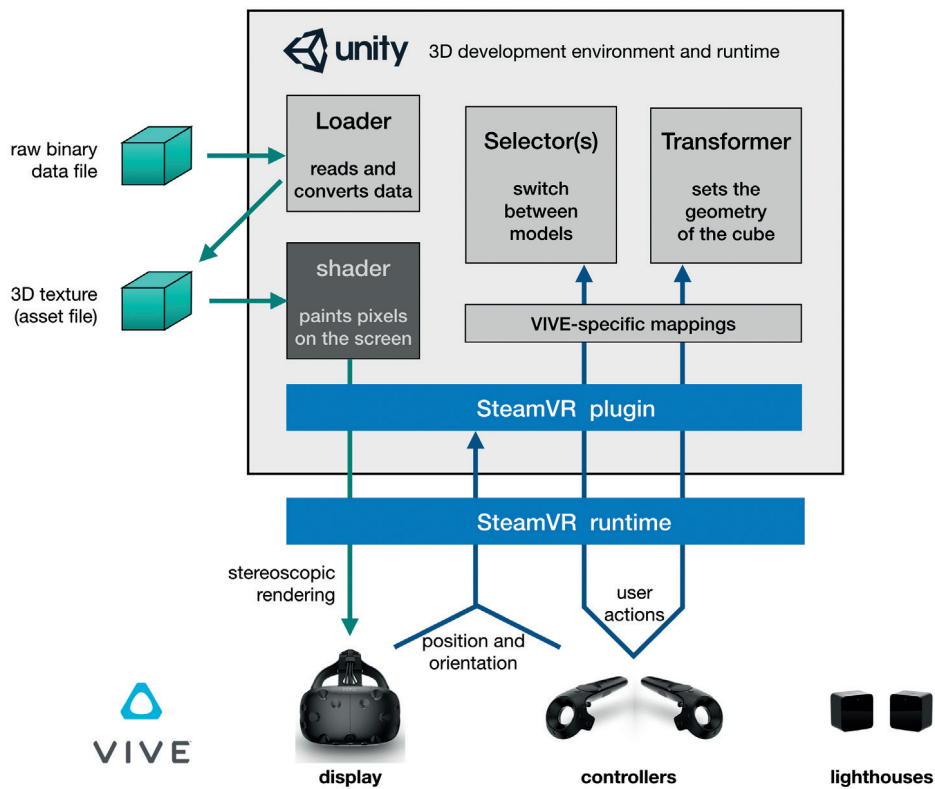

*Figure 4.* Sketch of the different software and hardware components. The grey boxes represent our scripts (written in C#, except the shader, which is written in Cg). The green arrows show the data flow when the full data cube is loaded for volume rendering. The blue boxes and arrows represent the SteamVR layer that interfaces Unity with VIVE; it takes care of stereoscopy and tracking and listens to user inputs. Credit: RIKEN/ABBL

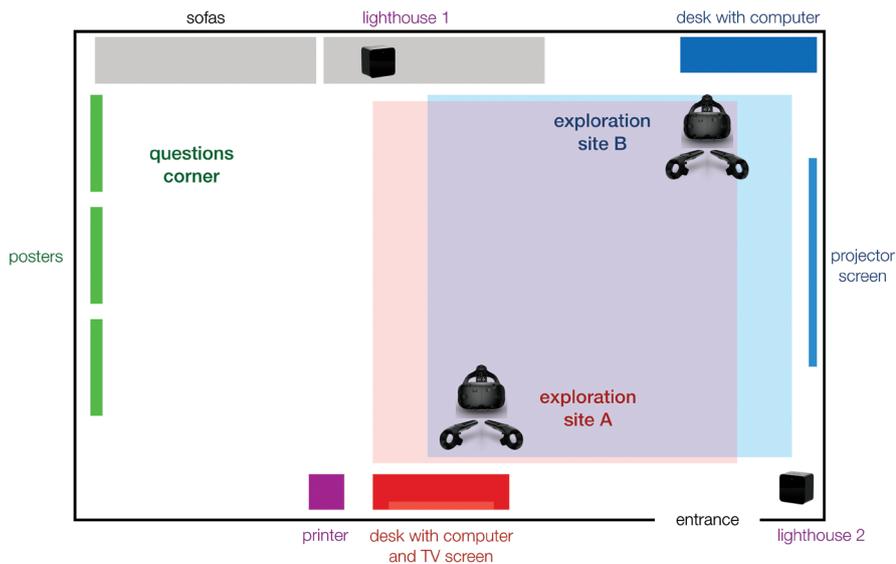

*Figure 5.* Sketch of the room layout. Two VR systems were installed (in red and blue) with overlapping exploration areas. Also in the room were a set of posters explaining the research conducted in the laboratory and a questions corner where visitors were free to ask about anything related to astronomy (both were accessible independently of the VR demo, so as not to rely solely on the VR hardware). Credit: RIKEN/ABBL

(currently available in English, Japanese and French).

## Points to Consider and Lessons Learnt

Setting up a public VR experience was a first for our laboratory and presented some new challenges.

### Using Space: Room Layout

The layout of the room assigned to our laboratory is shown in Figure 5 and in photographs in Figure 6. Two VIVE devices were set up at opposite corners of an area of approximately ten square metres, which allowed us to have two independent exploration sites in a limited space, while still allowing people to come and go in between. A nice feature of the VIVE's lighthouse system is that a single pair of devices can be used for tracking multiple headset/controller sets in the same room. The hardware setup worked well, considering how busy the room was at times, with very few technical glitches.

The live VR view was projected in two-dimensional (2D) form on a screen, so that others in the room could see what the person with the headset was doing; this was particularly useful so that groups of people could share the experience. This 2D preview did not, however, take away from the VR experience which was such a step up in terms of the level of immersion. It was still able to impress despite a 2D preview having been seen.

### Managing Time: Ticketing System

An important aspect to consider when showing such a VR demo to the public is that headsets offer a one-person-at-a-time experience. Based on the interest shown during the RIKEN Centennial Meeting, we knew that we needed a ticketing system in place for the RIKEN Open Day (and that we needed to announce this in the event programme). The day was split into six one-hour blocks, with six ten-minute slots per hour and with both VIVEs operating in parallel. This could in principle accommodate 72 guests. To avoid being sold out too early and to distribute tickets throughout the day, we gave out the twelve tickets for any hour block no more than one hour before the start of this block, that is, twelve new tickets became available for the following hour every hour.

We knew our VR booth would be popular, but it went well beyond our expectations. A long line formed in the corridor when the event opened. As a downside, ticketing proved to be the hardest task. Because so many people were lining up, we decided to accept small groups of people per ticket, as a trade-off between the number of people accommodated and the time spent by each person. The demo was given to approximately 160 people (more than double the expected headcount). To our surprise, many visitors were willing to wait in line for an hour until the next time tickets became available so they could be sure to secure one. This indicated that the potential of using VR technology to get the public interested in our research was massive and that in future, we need to be prepared for bigger crowds, with professional crowd management or a different way of distributing tickets. One solution would be to request online registration prior to the event. Another would be to use a lottery system or contest.

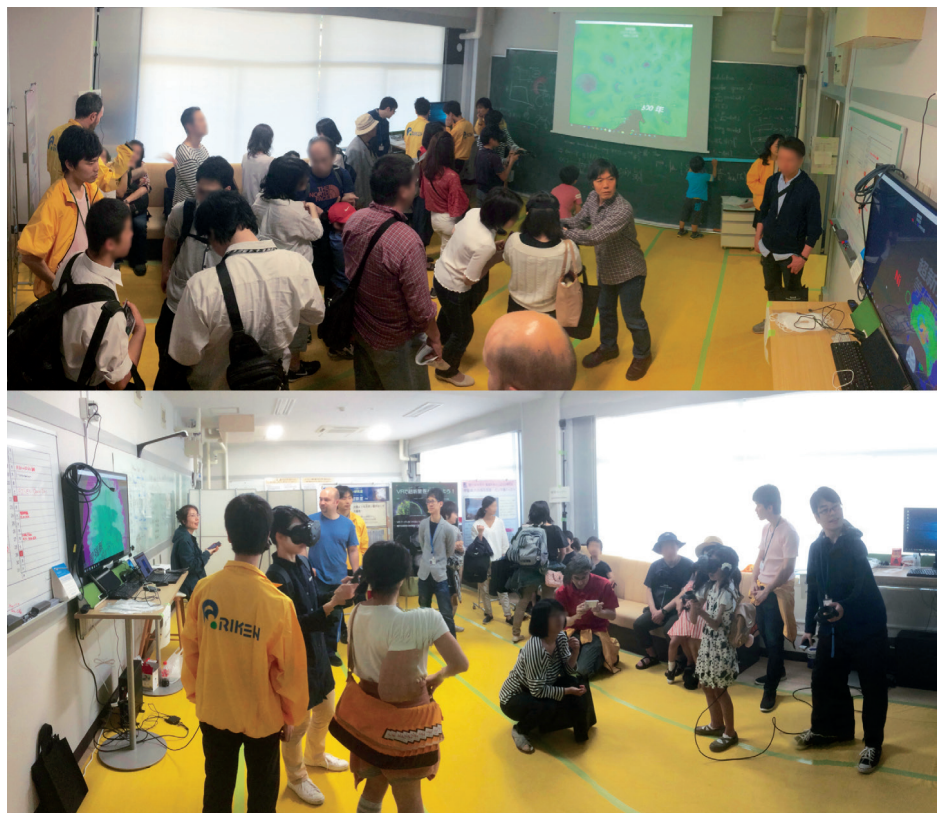

*Figure 6.* Panoramic photographs of the room during the event. Credit: RIKEN/ABBL

### Team Work: Role Assignment

Delivering an experience of this kind requires a coordinated team effort. We had identified the following roles prior to the event:

- A doorperson to welcome people, explain the workflow and distribute the tickets. The workload of this individual was underestimated. Because the VR booth was unusual, quite a lot of explanation was necessary. This role requires two people at the very least at any time.

- Navigators to give the demo (one per headset) and explain the science, how it was simulated and visualised and how users can interact with it. There was a sizable pool of seven Japanese-speaking navigators present all day (all but one was trained on the demo beforehand). This made it possible to rotate them in each hour block, which is good because continuously giving the demo can be exhausting.

- Photographers to take, print and hand out photos of the participants. Although many members were available for this, it is probably sufficient to have two per block.

### Briefing the Users

When ticket holders returned for their turn, they were handed two information sheets to read while waiting. One sheet lists basic precautions for using VR, which is a formal requirement, even though the risks are limited for our demo, and indeed, no one had any serious issues. The level of realism is such that some people were apparently genuinely scared when the controllers were first handed to them within the virtual world (others were just mesmerised). The other sheet shows the layout of the controllers, so that the people can familiarise themselves with the different buttons before putting on the headset. This was useful as it somewhat reduced the load on the navigators and saved time during the demos. Once the user is in VR, only verbal communication is possible.

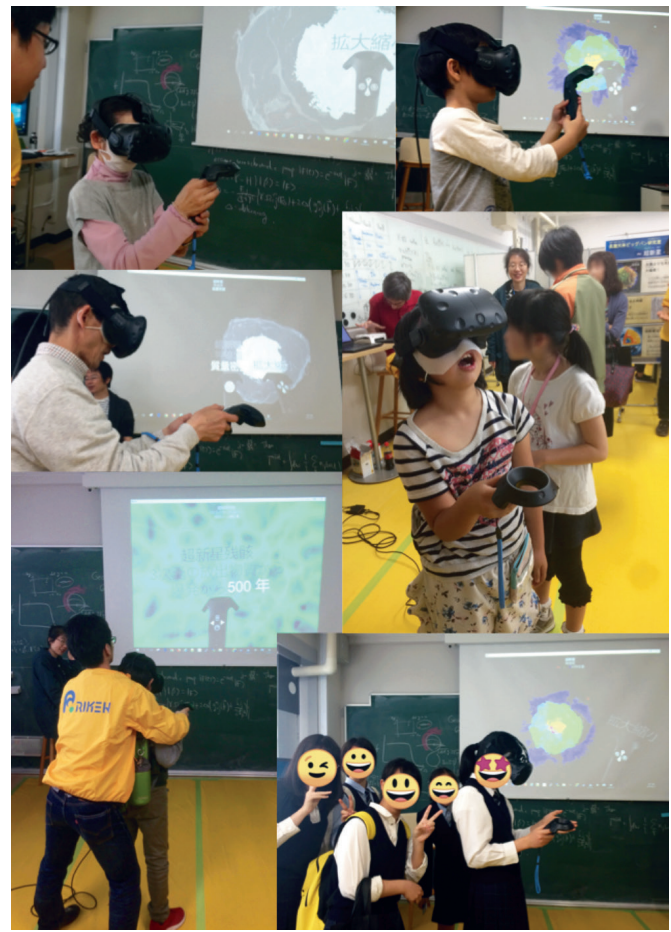

**Figure 7.** Photo collage of visitors experiencing the VR demo. Credit: RIKEN/ABBL

### Recording User Reactions

We had two methods of recording participants' reaction to the demo: the verbal account from the navigators and the comments written in the guest book, which the team tried to collect from each person. Visitors enjoyed the demo, despite the long wait; many were completely new to VR technology and so were very impressed by being able to hold the model in their hands or step in and peer inside, which has been described as feeling "like magic" the first time[20]. Rewardingly for us, people found the experience all the more remarkable since they were looking at actual astronomy data made by scientists: many expressed their gratitude that they could get to experience supernovae in such an intimate and personal way[21].

Given the short amount of time available for each participant, it was not possible to present all the aspects of the demo to everyone, and so the navigators focused on what generated the most enthusiasm. During such a large public event, visitors will always have different backgrounds and interests. So, in the future, we may build progressive disclosure in the software, with different user modes from beginner to advanced so that the demo can be better adjusted according to the user reactions.

**Collecting Statistics**

We were glad to have a varied crowd visiting our booth, see a sample in Figure 7. The age groups ranged from toddlers to the elderly (quite unexpectedly, both populations were able to use the headset, although the controllers proved more challenging), with most people coming in families, plus a number of groups of middle- and high-school students. A difference among genders was not apparent. We are not able to provide more quantitative demographic data for all the visitors. Given the nature of the event, we lacked both the time and the incentive to investigate this in detail, as visitors moved quickly and freely between the many booths at the Open Day event. Ideally, we would also like to know if they had experienced VR before (and in what context) and the extent to which the presence of a VR booth motivated them to visit us. How to obtain this information in the least disruptive way remains an open question.

We know that some people were genuinely interested in the science (for example, a woman who follows the institute's news on its YouTube channel), while others were coming to try the technology (like a student who clearly stated having an ambition to make a career in VR). The youngest participants simply enjoyed the aesthetics and interactivity and got a welcoming introduction to the research institute, which their parents also appreciated. It is now our task to capitalise on the novelty of the technology for the promotion of science. In the longer run, we will try to assess the impact of the use of VR on the interest visitors show in our research and their perception of our laboratory.

**Creating a Connection: Souvenir Photos**

The VR demo is an intense, but short experience. In the hope of creating a lasting link between the visitors and our laboratory, the team leader had the idea to make and give out postcards of each participant trying the demo[22]. The laboratory invested in a small photo printer (similar to the ones found in photo booths), and the photographers used their smartphones to take snapshots and upload them on a shared online folder, so they could quickly be printed on site. We were able to give everyone (provided they approved it) a souvenir of their unique experience, branded with the RIKEN logo and our laboratory's name (a re-created example is shown in Figure 8). This proved very successful, and we highly recommend this practice. We are also re-printing a large collage of the photos to display on the wall at the institute to promote our activities.

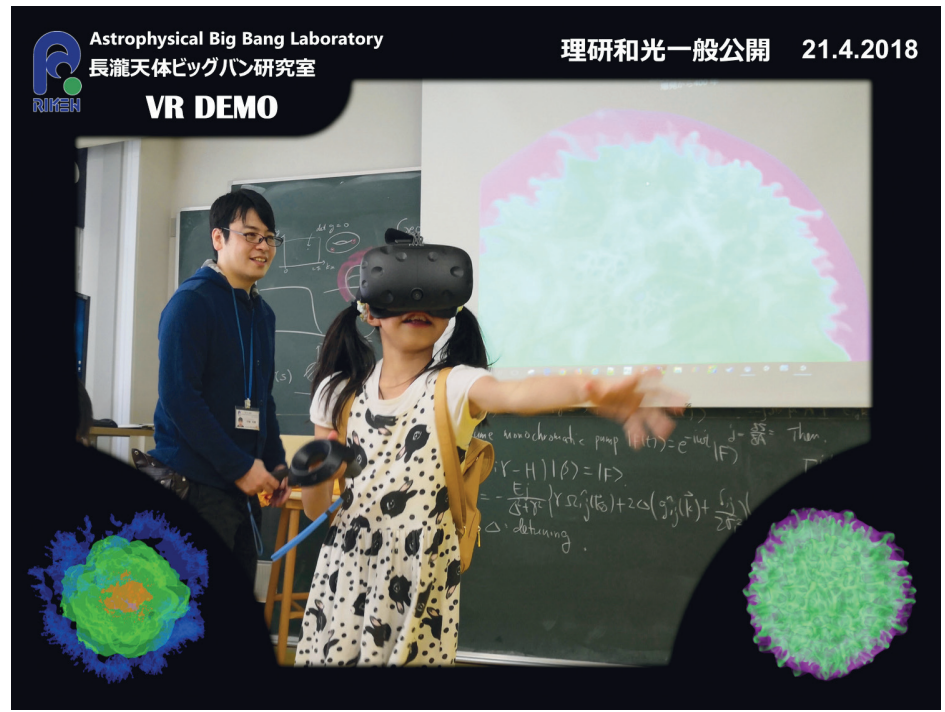

*Figure 8. Example of a souvenir photo that was printed and given during the event. Credit: RIKEN/ABBL*

**Perspectives**

We believe that our experiment was successful and shows the potential impact of 3D immersive visualisation in driving public interest in astrophysics research. The ABBL will continue to use VR to promote the work conducted at the laboratory. The VR booth will return for the next RIKEN Open Day, and we are contemplating the possibility of setting up a (semi-)permanent exhibition. Since several VR arcades have recently opened in Tokyo, why not use them to offer a science-based experience?

Most astronomers have not made the transition to the kinds of tools we have described here, chiefly because of the high costs and the difficulty (real or perceived) of use. To our knowledge, the only other application of modern VR for astronomy data is visualisation of the 3D structure of the SNR Cassiopeia A, as reconstructed from observations, produced by NASA's Chandra team[23]. However, recent developments on both the hardware and software sides have considerably lowered the cost of entry for VR, and we argue that now is a good time to get started — for research as well as outreach. It is also the time to build a community for VR astronomy and generally, for advanced visualisation in science. By sharing our method and insights[24], we hope to help others create innovative and engaging experiences for the public. Our generation may not be able to actually travel to the stars, but we can certainly do so virtually.

**Acknowledgments**

First, we warmly thank the team that made it possible to deliver this experience at the RIKEN Open Day: Shigehiro Nagataki and Tamaki Shibasaki (ABBL chief scientist and secretary); navigators Masaomi Ono, Hirotaka Ito and Susumu Inoue (post-doctoral researchers at RIKEN); students from Rikkyo University Masanori Arakawa, Hiroyoshi Iwasaki, Atsuhiro Ebata and Yutaro Yoshino; and photographers Haoning He, Oliver Just, Donald Warren and Gilles Ferrand (post-doctoral researchers at RIKEN). Extra thanks are due to Masaomi Ono for the Japanese translation.

We also acknowledge our colleague Fritz Röpke (Heidelberg Institute for Theoretical Studies and Universität Heidelberg) for providing us data from the supernova simulations as part of Ferrand's research project.

Ferrand extends special thanks to Jayanne English (University of Manitoba) for prompting his initial foray into VR and for remotely testing the early SN demo before the ABBL acquired its own headset.

## Notes

[1] More information on Astrophysical Big Bang Laboratory: nagataki-lab.riken.jp/research_en.html

[2] More information on RIKEN (理研), the largest public research institute of Japan: www.riken.jp/en/about/

[3] In augmented reality (AR), it is possible to place computer-generated elements in the real world. Our project however does not use this technique.

[4] During the international workshop "Theories of International Big Bangs" held on our campus (2017-11-06 to 2017-11-17), the author included a live demonstration as part of his talk, which served both as an introduction to the scientific use of VR and as an illustration of his research work.

[5] Open day information: http://openday.riken.jp (the supernova experience in VR is announced on page 10 of the PDF pamphlet).

[6] A supernova (SN) marks the end point of the evolution of some stars (either a massive star or a white dwarf). As the ejected stellar material rushes away from the explosion centre, it triggers a strong blast wave that ionises and energises the ambient matter. It is this interaction between the ejecta and the interstellar medium (ISM) that produces the supernova remnant (SNR). While the SN phase may be measured in seconds, the SNR phase may last for thousands or tens of thousands of years, until the remnant dissolves into the ISM.

[7] The pilot project, conducted by G. F. at the University of Manitoba, was a collaboration between the departments of Physics and Astronomy (principle investigator Jayanne English, radio astronomer) and Computer Science (principle investigator Pourang Irani, head of the Human Computer Interaction lab): hci.cs.umanitoba.ca/projects-and-research/details/3d-visualization-of-astronomical-data-using-immersive-displays

[8] Google Cardboard: vr.google.com/cardboard/

[9] Visbox CAVE: www.visbox.com/products/cave/

[10] VIVE: www.vive.com

[11] Unity can be downloaded from unity3d.com. It is commercial software, although it is free to use for developers with our needs and usage. It is available on all platforms, although at this time, the VIVE hardware is officially supported only on Windows.

[12] Another generic software for 3D modelling that has proved popular for scientific applications is Blender, although it has been used more often for off-line rendering rather than for interactive visualisation. The most relevant programme for what we are doing is the FRELLED software by Rhys Taylor (www.rhysy.net/frelled-1.html).

[13] An example in the field of oceanography and meteorology is TerraViz, the visualisation component of NOAA's Earth Information System, which is entirely built using Unity, described as "an ideal choice for providing a wealth of data to a user in real-time". www.esrl.noaa.gov/neis/library/terraviz-video.html

[14] Another software for meteorologists, MEVA, combines "the power and variety of data abstraction methods of a visualisation tool (ParaView) with the interaction and presentation capabilities offered by computer-games engines (Unity)". www.ufz.de/index.php?en=37915

[15] Steam VR: store.steampowered.com/steamvr

[16] Our software is still very much a prototype. It is not yet a complete visualisation package for astronomy; rather, it is a workbench for exploration. For a full-fledged solution for scientific visualisation in VR, in a CAVE environment, see the VFIVE application originally developed by Akira Kageyama.

[17] Visualisation software VisIt: wci.llnl.gov/codes/visit/

[18] Visualisation software Python/scikit-image: scikit-image.org

[19] In VR, real-time rendering is critical because the user determines the current viewpoint at any time. A noticeable delay between when they move and when the environment responds breaks the 3D illusion and can cause significant discomfort.

[20] This was expressed by a number of すごい (sugoi) interjections, which loosely means "terrific" in Japanese and is the idiomatic equivalent of "awesome".

[21] One eleven year old even reported feeling the "wind" from the explosion during the demo… even though we have not deployed such multi-sensory haptics yet!

[22] The next step will be to show the person interacting with the virtual model by compositing the real-world photograph with the computer-rendered display.

[23] Project "Walking among the stars", led by Kimberly Arcand (chandra.harvard.edu/vr/)

[24] We are considering options to share the Unity project with colleagues and to share the final product with the public. If you are interested in trying our demo and have the hardware, please contact us directly.

### Biographies

**Gilles Ferrand** is an astrophysicist at the Astrophysical Big Bang Laboratory (ABBL) of RIKEN. He is working mainly on particle acceleration and supernova remnants. He also has a keen interest in scientific visualisation.

**Don Warren** is an astrophysicist for the Interdisciplinary Theoretical and Mathematical Sciences (iTHEMS) Program at RIKEN. He also helps organise and run the monthly outreach event Nerd Nite Tokyo.